\RequirePackage{bm}
\documentclass[journal=jctcce,manuscript=article,layout=onecolumn,sections=on]{achemso}

\usepackage[version=3]{mhchem} 

\usepackage[version=3]{mhchem} 
\usepackage{amssymb,amsmath}
\usepackage{bm}
\usepackage{amsfonts}
\usepackage{braket}
\usepackage{natbib}
\usepackage{appendix}
\usepackage{graphicx}
\usepackage{epstopdf}
\usepackage{threeparttable}
\usepackage{booktabs}
\usepackage{lscape}
\usepackage{color}
\usepackage{soul}
\usepackage{multirow}
\usepackage{setspace}
\usepackage{multirow}
\usepackage{breqn}
\usepackage{bm}
\usepackage{siunitx}
\usepackage{geometry}
\usepackage{fancyref}
\usepackage[framemethod=tikz]{mdframed}

\author{GiovanniMaria Piccini}
\affiliation{Department of Chemistry and Applied Biosciences, ETH Zurich, c/o USI Campus, Via Giuseppe Buffi 13, CH-6900, Lugano, Ticino, Switzerland}
\altaffiliation{Facolt{\'a} di Informatica, Instituto di Scienze Computationali, Universit{'a} della Svizzera italiana (USI), Via Giuseppe Buffi 13, CH-6900, Lugano, Ticino, Switzerland}

\author{Dan Mendels}
\affiliation{Department of Chemistry and Applied Biosciences, ETH Zurich, c/o USI Campus, Via Giuseppe Buffi 13, CH-6900, Lugano, Ticino, Switzerland}
\altaffiliation{Facolt{\'a} di Informatica, Instituto di Scienze Computationali, Universit{'a} della Svizzera italiana (USI), Via Giuseppe Buffi 13, CH-6900, Lugano, Ticino, Switzerland}

\author{Michele Parrinello}
\email{parrinello@phys.chem.ethz.ch}
\affiliation{Department of Chemistry and Applied Biosciences, ETH Zurich, c/o USI Campus, Via Giuseppe Buffi 13, CH-6900, Lugano, Ticino, Switzerland}
\altaffiliation{Facolt{\'a} di Informatica, Instituto di Scienze Computationali, Universit{'a} della Svizzera italiana (USI), Via Giuseppe Buffi 13, CH-6900, Lugano, Ticino, Switzerland}

\title[\texttt{achemso} demonstration]
{Metadynamics with Discriminants: a Tool for Understanding Chemistry}

\begin{document}


\begin{abstract}
\noindent

We introduce an extension of a recently published method\cite{Mendels2018} to obtain low-dimensional collective variables for studying multiple states free energy processes in chemical reactions.
The only information needed is a collection of simple statistics of the equilibrium properties of the reactants and product states.
No information on the reaction mechanism has to be given.
The method allows studying a large variety of chemical reactivity problems including multiple reaction pathways, isomerization, stereo- and regiospecificity.
We applied the method to two fundamental organic chemical reactions.
First we study the \ce{S_N2} nucleophilic substitution reaction of a \ce{Cl} in \ce{CH_2 Cl_2} leading to an understanding of the kinetic origin of the chirality inversion in such processes.
Subsequently, we tackle the problem of regioselectivity in the hydrobromination of propene revealing that the nature of empirical observations such as the Markovinikov's rules lies in the chemical kinetics rather than the thermodynamic stability of the products.
\end{abstract}


\section*{Introduction}
Molecular dynamics (MD) is a powerful tool for studying several properties of different chemical systems.
However, chemistry is mostly interested in activated processes that brings reactants to products, namely chemical reactions.
Thermodynamically speaking, reactants and products are free energy basins separated by a high barrier.
In most cases, such barriers are associated to extremely long time scales that are incomparably larger than those reachable in MD.
For many years this has limited the application of MD to problems distant from the main interest of chemistry.

Since the landmark paper of Torrie and Valleau\citep{Torrie1977} in which umbrella sampling was introduced, a large variety of methods has been proposed to study rare events.
One of the most popular and widely applied method is metadynamics\citep{laio_parrinello_2002,Barducci2008}.
In this method a history dependent adaptive bias is recursively added to the underlying free energy of the system.
Its effect is to enhance the fluctuations in the basins accelerating the transitions between reactants and products.

Umbrella sampling and Metadynamics, like other methods, rely on the definition of some order parameters, or collective variables (CVs), able to \emph{discriminate} the thermodynamics states of interest in the free energy space.
The CVs are expected to capture the slowest motions of the system, i.e. those degrees of freedom whose sampling needs to be accelerated\citep{valsson_2016}.
More formally, a CV is a function $s(\mathbf{R})$ of the microscopic atomic coordinates $\mathbf{R}$.
Over the years a large variety of CVs has been proposed to enhance the sampling for different free energy processes ranging from nucleation problems\cite{Piaggi2017} to protein folding\citep{schaffer_parrinello_2016}.

From a simple point of view, chemical reaction can be seen as a drastic bonds rearrangement.
Reactants and products can be discriminated by their bonding topology.
After the system has reacted some bonds may have formed while others may have been broken.
A good CV should preferably combine all this information into the most simple mathematical expression.
By means of chemical intuition one can select a set of simple bond distances and combine them into one single linear combination.
For simple cases this may be an easy task.
However, if the reaction is somewhat complex an a priori evaluation of the coefficients of the linear combination may lead to a very poor convergence or, even worse, to a wrong description of the physics.
A further complication occurs if the reactant state can evolve into different products or if the latter can isomerize into one another.
In a recent publication by our group\citep{Mendels2018} we have shown that part of the chemical intuition process can be automatized by means of a method that we called harmonic linear discriminant anlysis (HLDA).
In this work we extend this method to a multi class (MC) problem in order to treat multiple thermodynamic states simultaneously.
We shall refer to this extension as MC-HLDA.

The paper starts with an introduction to MC-HLDA.
Two examples have been considered to elucidate the power of the method in understanding fundamental chemical processes.
The first is the \ce{S_N2} nucleophilic substitution of a chlorine atom in dichloromethane while the second is the electrophilic addition of hydrogen bromide the propene.

\section*{Method}
Even for rather small systems a chemical reaction may involve several slow degrees of freedom that play an important role in determining the transition between reactants to products.
However, enhanced sampling methods like metadynamics tend to converge slowly when a large number of CVs is employed and the interpretation of the result may lack of sufficient clarity.
Our goal is to pack large sets of descriptors into few linear combinations that can still describe accurately the physics of the problem.

We assume that we are dealing with a chemical reaction characterized by $M$ states separated by high free energy barriers, e.g. reactants that may evolve into different products, and that $N_d$ descriptors $d(\mathbf{R})$ suffice to characterize and distinguish them like distances between pairs of atoms involved in breaking and/or forming a chemical bond.
Since transitions between the $M$ states are rare we can compute for each $I$-th state the expectation value $\boldsymbol{\mu}_{I}$ of the $N_d$ descriptors, a vector collecting the averages of each $d(\mathbf{R})$ for state $I$, and the associated multivariate variance $\boldsymbol{\Sigma}_{I}$, i.e. the covariance matrix.
These quantities can be straightforwardly evaluated in short unbiased runs. 
Our goal is to find an $M-1$ dimensional projection of these $M$ sets of data along which their mutual overlap is minimal. 
Multi class linear discriminant analysis aims at finding the directions along which the projected data are at best separated.
To do this one needs a measure of their degree of separation. 
Following up our previous work\citep{Mendels2018} we introduce the Fisher's\citep{Fisher1936} criterion. This is given by the ratio between the so called ``between class'' $\mathbf{S_b}$ and the ``within class'' $\mathbf{S_w}$ scatter matrices transformed by the rotation matrix $\mathbf{W}$.  
The former is measured by the square of the distance  between the projected means

\begin{equation}
\label{mean_transf_mat}
\mathbf{W}^T \mathbf{S}_b \mathbf{W}
\end{equation}

\noindent
with

\begin{equation}
\label{between_class}
\mathbf{S}_b = \sum_I^M \left( \boldsymbol{\mu}_I - \boldsymbol{\overline{\mu}} \right)\left(  \boldsymbol{\overline{\mu}} - \boldsymbol{\mu}_I \right)^T
\end{equation}

\noindent
where $\boldsymbol{\mu}_I$ is the mean value of the $I$-th class and $\boldsymbol{\overline{\mu}}$ is the overall mean of the datasets, i.e. $\boldsymbol{\overline{\mu}} = 1/M \sum_I^M \boldsymbol{\mu}_I$.

\noindent
The within scatter is the pooled covariance leading to the expression 

\begin{equation}
\label{cov_transf_mat}
\mathbf{W}^T \mathbf{S}_w \mathbf{W}
\end{equation}

\noindent
with 

\begin{equation}
\label{within_class}
 \mathbf{S}_w = \sum_I^M \boldsymbol{\Sigma}_I
\end{equation}

\noindent
being $\boldsymbol{\Sigma}_I$ the covariance matrix of state $I$.
The Fisher’s object function\citep{Fisher1936} then reads like a Rayleigh ratio
 
\begin{equation}
\label{fisher_ration}
\mathcal{J(\mathbf{W})} = \frac{\mathbf{W}^T \mathbf{S}_b \mathbf{W}}{\mathbf{W}^T \mathbf{S}_w \mathbf{W}}.
\end{equation}

\noindent
We want $\mathbf{W}^T \mathbf{S}_b \mathbf{W}$ to be maximized subject to the condition $\mathbf{W}^T \mathbf{S}_w \mathbf{W}=1$, corresponding to Lagrangian,

\begin{equation}
\label{maximizer}
\mathcal{L}_{LDA} = \mathbf{W}^T \mathbf{S}_b \mathbf{W} + \lambda \left( \mathbf{W}^T \mathbf{S}_w \mathbf{W} - 1 \right).
\end{equation}

\noindent
This problem can be solved by means of a generalized eigenvalue solution of the form

\begin{equation}
\label{gen_eig}
\mathbf{S}_b \mathbf{W} = \lambda \mathbf{S}_w \mathbf{W}.
\end{equation}

\noindent
Assuming that the inverse of $\mathbf{S}_w$ exists the above equation can be reduced to the standard eigenvalue problem in the form of

\begin{equation}
\label{std_eig}
\mathbf{S}_w^{-1} \mathbf{S}_b \mathbf{W} = \lambda \mathbf{W}
\end{equation}

\noindent
The theory of linear discriminants tells us that for $M$ classes we get $M-1$ non-zero eigenvalues.
The eigenvectors associated to these eigenvalues correspond to the directions in the $N_d$ space along which the distributions of the $M$ states exhibit the least overlap.
Thus, they can be used as CVs in metadynamics or other CV based enhance sampling methods.

As discussed in our previous work\cite{Mendels2018}, from the rare event point it is more appropriate to give more weight to the state with the smaller fluctuations.
In other words focus more on those states that are more difficult to get into and escape from. 
For these reasons we have proposed a different measure of the scatter and rather than using the arithmetic average we base the measure of the within scatter matrix on the harmonic average as follows:

\begin{equation}
\label{harmonic_mean}
\mathbf{S}_w = \frac{1}{\frac{1}{\boldsymbol{\Sigma}_A} + \frac{1}{\boldsymbol{\Sigma}_B} + \dots + \frac{1}{\boldsymbol{\Sigma}_M}}.
\end{equation}

\noindent
As discussed in our previous work, from a machine learning point of view the more compact states are better defined and thus have a larger weight in determining the discriminants.
Experience has shown us that this second choice is far superior.

\section*{Results and Discussion}

\subsection*{Nucleophilic substitution of a \ce{Cl} atom in \ce{CH_2 Cl_2}}

\begin{figure}
	\includegraphics[width=0.5\columnwidth]{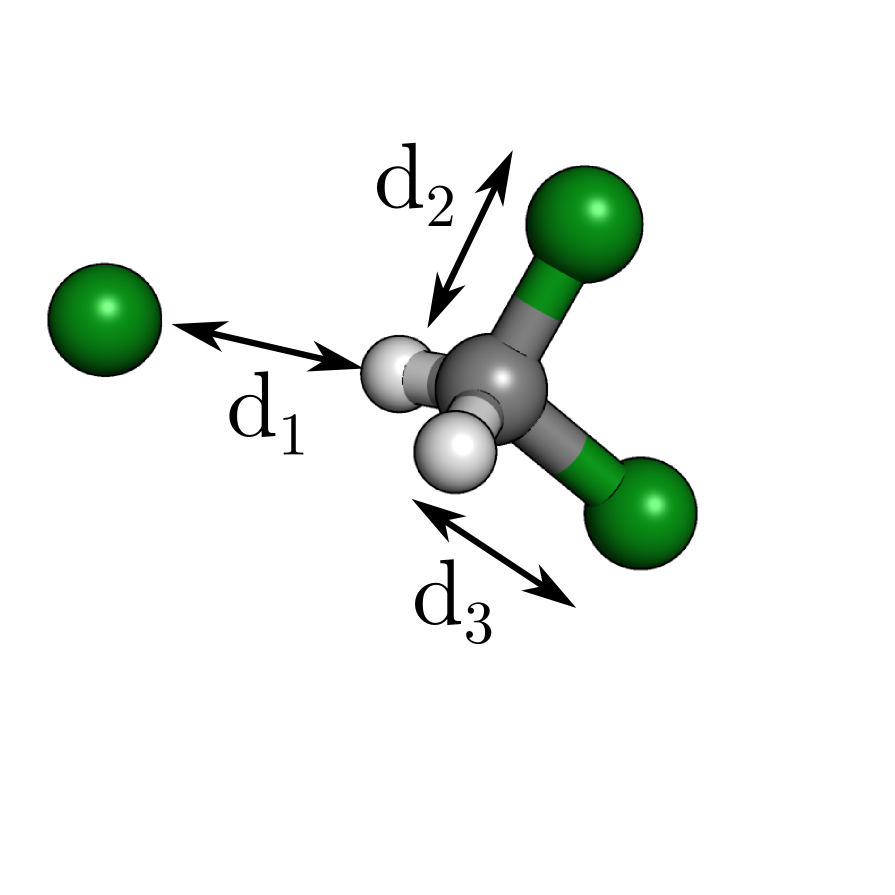}
   	 \caption{Carbon-chlorine distances used as descriptors for the \ce{S_N2} reaction.}
    	\label{fig:dist_sn2}
\end{figure}

In order to illustrate the power of MC-HLDA in the study of chemical reactions we start with a simple \ce{S_N 2} nucleophilic substitution, namely the substitution of a chlorine atom in the compound \ce{CH_2 Cl_2} by a \ce{Cl^-} anion.
For this system reactants and products are formally identical and the statistics of the state descriptors in the free-energy basins is permutationally equivalent.
To enhance sampling between the three basins we employed the distances between the chlorine atoms and the central carbon atom similarly to what done recently by Pfaender et al. \cite{fleming_pfaender_2016} and by our group \cite{piccini2016} (see Fig. \ref{fig:dist_sn2}). 
Ab initio MD simulations at 300 K were performed using the CP2K\cite{hutter_vandevondele_2014} package patched with the PLUMED2\cite{tribello_2014} code.
A time step of 0.5 fs was used.
The PM6 Hamiltonian was used to calculate energies and gradients.
To control the temperature the systems has been coupled to the velocity rescaling thermostat of Bussi et al.\cite{bussi_parrinello_2007} every 100 MD steps.
Here, the thermostat aims at mimicking in a simple manner the effect of the environment as we do not intend to simulate a process in the gas phase for which a microcanonical simulation must be considered.
Well-tempered Matadynamics has been used to enhance the sampling along the MC-HLDA CVs for a total simulation time of 3 ns.
For further computational details we refer the interested reader to the Supporting Information.

\begin{figure}
	\includegraphics{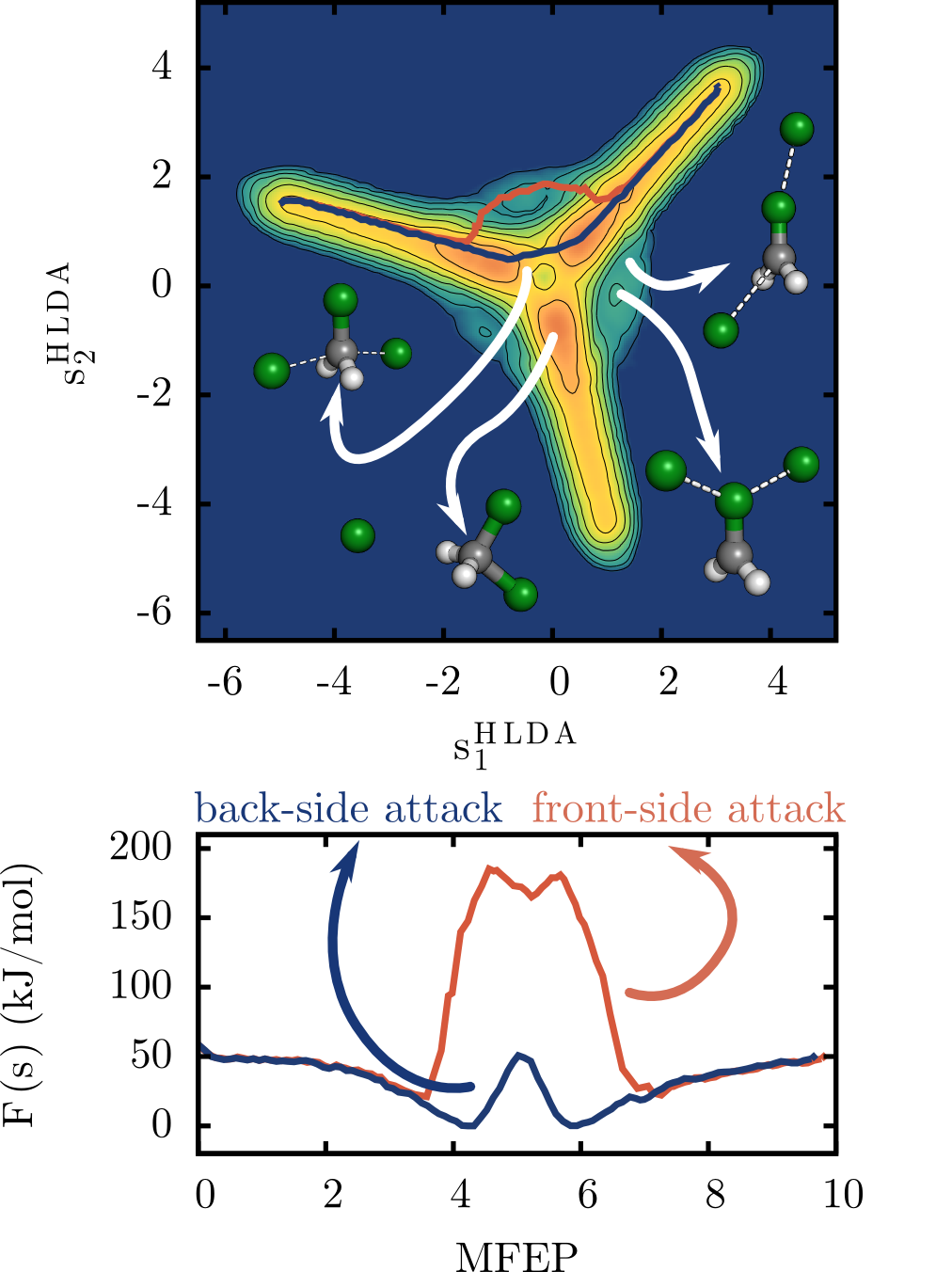}
   	 \caption{Upper panel: free energy surface for the nucloephilic substitution of a chlorine atom in dichloromethane showing the two possible reaction paths and associated relevant reference structures. Lower panel: minimum free energy paths along for the two possible mechanisms of the reaction, namely the low-barrier back-side attack (blue line), and the high-barrier front-side attack (orange line).}
    	\label{fig:fes_sn2}
\end{figure}

These three descriptors represent a good choice in describing the slow motions of the atoms in going from reactants to products.
More precisely, they are related to the slowest collective oscillations, i.e. low vibrational frequencies, as they involve the dynamics of the heaviest atoms of the system\cite{hase1990,truhlar1989}.
In order to calculate the three $\boldsymbol{\mu}_{I}$ and $\boldsymbol{\Sigma}_{I}$ needed to solve eq. \ref{std_eig} we just have to perform the calculation in one of the basins and use symmetry to obtain the other two sets of data.
Since this is a 3-class problem the dimensionality reduction operated by MC-HLDA provides two linear combinations of  descriptors.
Hence, the CV space is two dimensional.
Fig. \ref{fig:fes_sn2} reports the free energy surface (FES) obtained by enhancing the sampling along the two CVs by means of metadynamics.
The variables provide an excellent discrimination of the states neatly reflecting the 3-fold symmetry of the problem.
Convergence of the FES with respect to simulation time and relative error bars estimated via bootstrap analysis can be found in the supporting information.

To extract further information we calculated the free energy profile along the minimum free energy path (MFEP)\cite{branduardi2013,ensing_parrinello_2004}.
The latter is obtained by taking a series of points lying along a guess path and minimized according to the nudged elastic band algorithm (NEB)\cite{neb1998}.
In principle such a MFEP is not required to go necessarily through the transition states.
However, in the present case it does and the conformations extracted from the apparent transition state do correspond to the classical back-side nucleophilic attack as depicted in blue in Fig. \ref{fig:fes_sn2}.
This is a clear indication of the quality of the CVs.
The estimated barrier along this path is about 50 \si{kJ/mol}.
The back-side attack is the most probable route in a \ce{S_N2} reaction mechanism and it would be responsible for the chiral center inversion if any chirality were present.
However, a closer inspection of the FES revealed the existence of another reaction pathway much higher in terms of free energy barriers.
The MFEP analysis shows that along this possible, although much less probable path, the system must overcome a barrier in the order of 200 \si{kJ/mol}.
Rather surprisingly, the analysis revealed the presence of a high energy intermediate in which the nucleophile and the leaving group coordinate the \ce{Cl^-} atom bonded to the central carbon atom.
In physical organic chemistry this mechanism is usually referred to as front-side attack and it is known experimentally\cite{Dougherty1974} and theoretically\cite{Schlegel1977,szabo_gabor_2015,hase2016} to be high in terms of energy barriers, thus, unlikely to happen.
In fact, this mechanism is expected to retain the chirality of the reactants a feature that is almost never observed in pure \ce{S_N2} reactions.
Therefore, it is clear that the nature of the \ce{S_N2} chirality inversion is a consequence of the chemical kinetics as the barrier for the back-side attack mechanism is much lower than the front-side one.

As a general and final remark in this section we would like to stress once more that no information on the route the system can take to go from reactants to products has been given as an input.
All this wealth of information on the system was hidden in the simple statistics that one collects from a short monitoring of the local fluctuations in the free energy basins.

\subsection*{Electrophilic addition of \ce{HBr} to propene}
\begin{figure}
	\includegraphics[width=0.5\columnwidth]{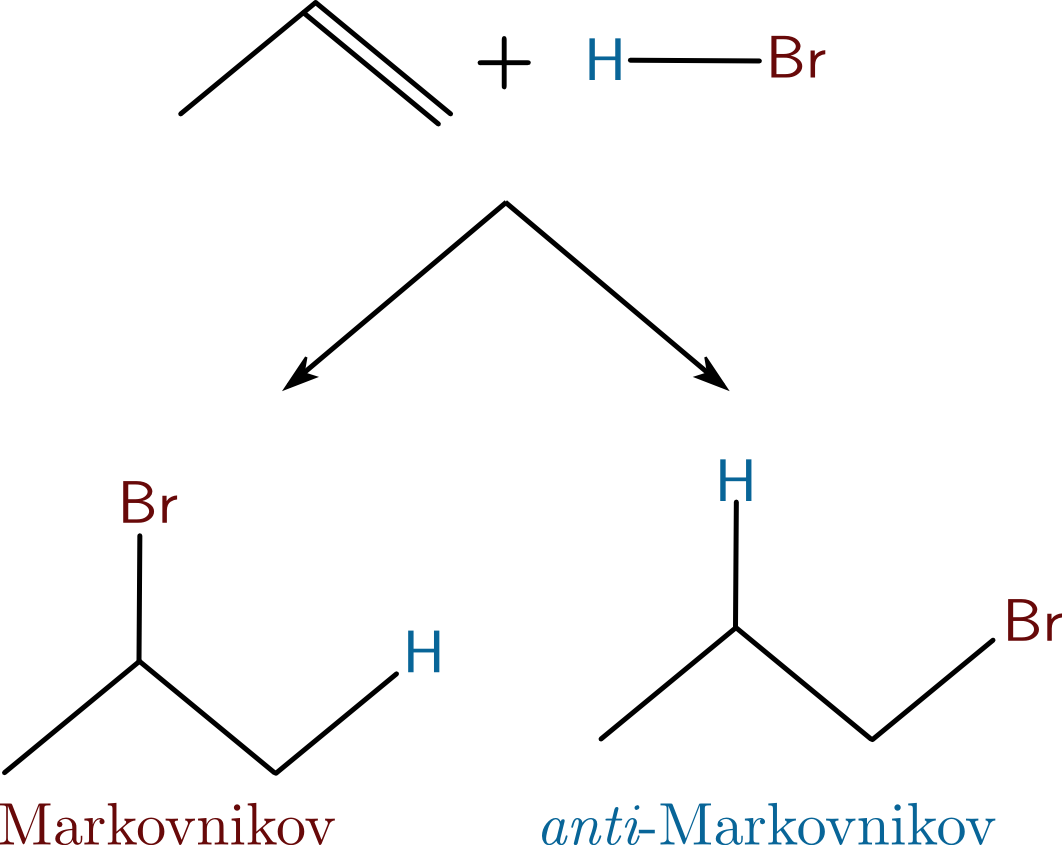}
   	 \caption{Reaction scheme for the hydrobromination of propene showing the two possible isomers classified as Markovnikov and \emph{anti}-Markovnikov products.}
    	\label{fig:react_es}
\end{figure}

Another fundamental type of organic reaction is the electrophilic addition.
One such process is the hydrobromination of propene.
Here, a hydrogen bromide molecule is used to break the propene double bond by adding the hydrogen atom and the bromine to the carbon atoms involved in the double bond.
Such a reaction can give rise to two different isomers depending on which carbon atoms the halide group will bind to (see Fig. \ref{fig:react_es}).

It is known that the halide atom prefers to bond to the most substituted carbon.
This because the mechanisms starts with the abstraction of the acidic hydrogen by a carbon atom of the double bond implying the formation of a carbocation.
Carbocations are much more stable if the carbon center is surrounded by other carbon groups rather than by hydrogen atoms.
In the case of propene the central carbon atom has a hydrogen atom formally substituted by a methyl group.
Therefore, this stabilizes the transient carbocation with respect to the less substituted carbon in the dobule bond.
The carbocation is formed in the transition region and is unstable, thus, the preference of the bromine atom to sit in the most substituted carbon has to be interpreted as a kinetic rather that a thermodynamic effect.
These  are to the so called Markovnikov rules\cite{Svensson1997,Aizman2002,Yang2008} and the associated reaction outcomes are referred as Markovnikov and anti-Markovnikov products (see Fig. \ref{fig:react_es}).

\begin{figure}
	\includegraphics[width=0.5\columnwidth]{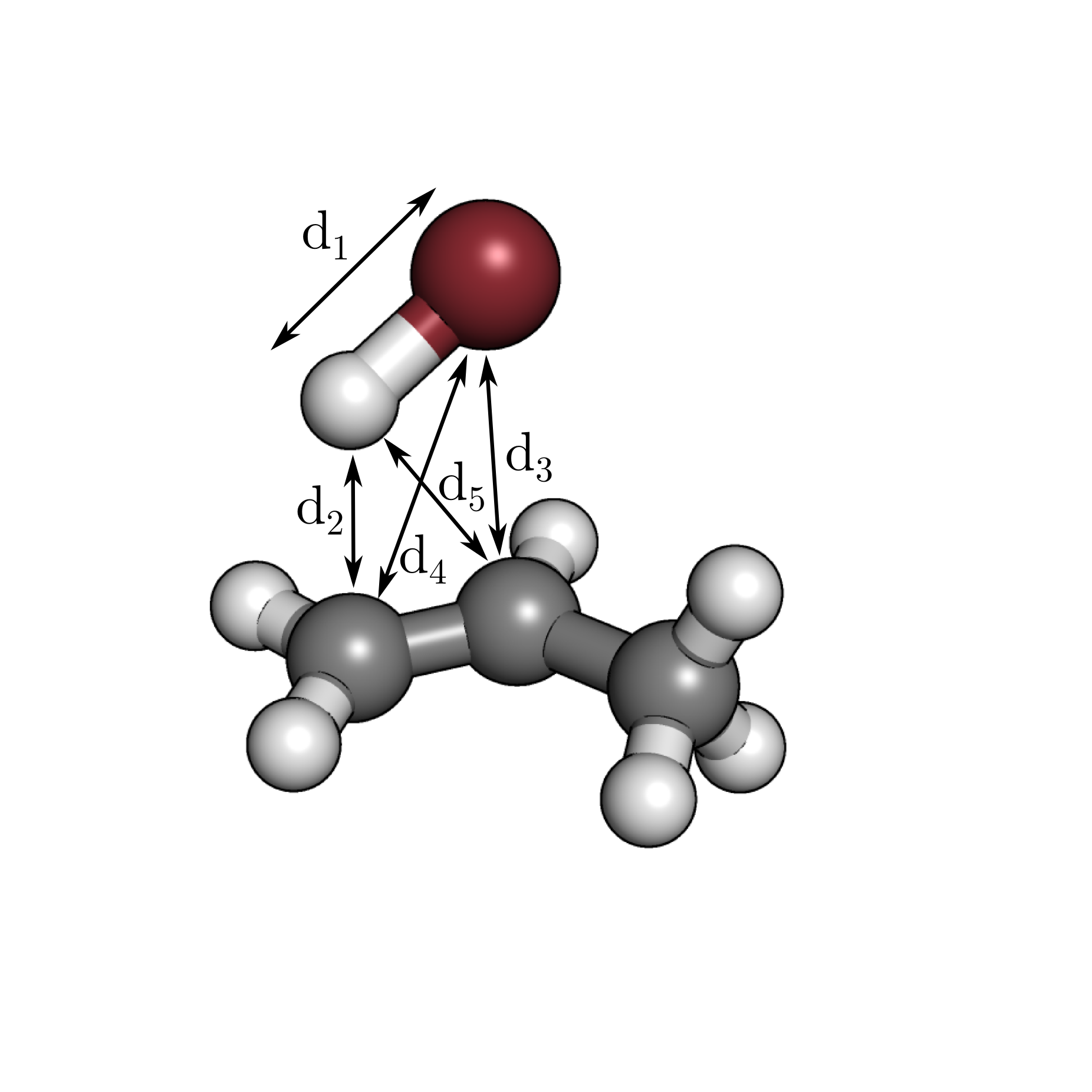}
   	 \caption{Fundamental distances used as descriptors for the hydrobromination of propene.}
    	\label{fig:dist_es}
\end{figure}

To study this particular reaction we applied MC-HLDA to the three states associated to reactants, Markovnikov, and anti-Markovnikov products states.
We used as descriptors the five distances illustrated in Fig. \ref{fig:dist_es}.
These CVs are both able to break and form the desired bonds but at the same time they embody the necessary information to discriminate between Markovnikov and anti-Markovnikov products.
Similarly to the previous case, ab initio MD simulations at 300 K were performed using the CP2K\cite{hutter_vandevondele_2014} package patched with the PLUMED2\cite{tribello_2014} code.
A time step of 1.0 fs was used.
The PM6 Hamiltonian was used to calculate energies and gradients.
To control the temperature the systems has been coupled to the velocity rescaling thermostat of Bussi et al.\cite{bussi_parrinello_2007} every 100 MD steps.
Again, the thermostat here aims at mimicking in a simple manner the effect of the environment.
Well-tempered Matadynamics using 3 multiple-walkers has been used to enhance the sampling along the MC-HLDA CVs for a total simulation time of 2.5 ns.
For further computational details we refer the interested reader to the Supporting Information.

\begin{figure}
	\includegraphics{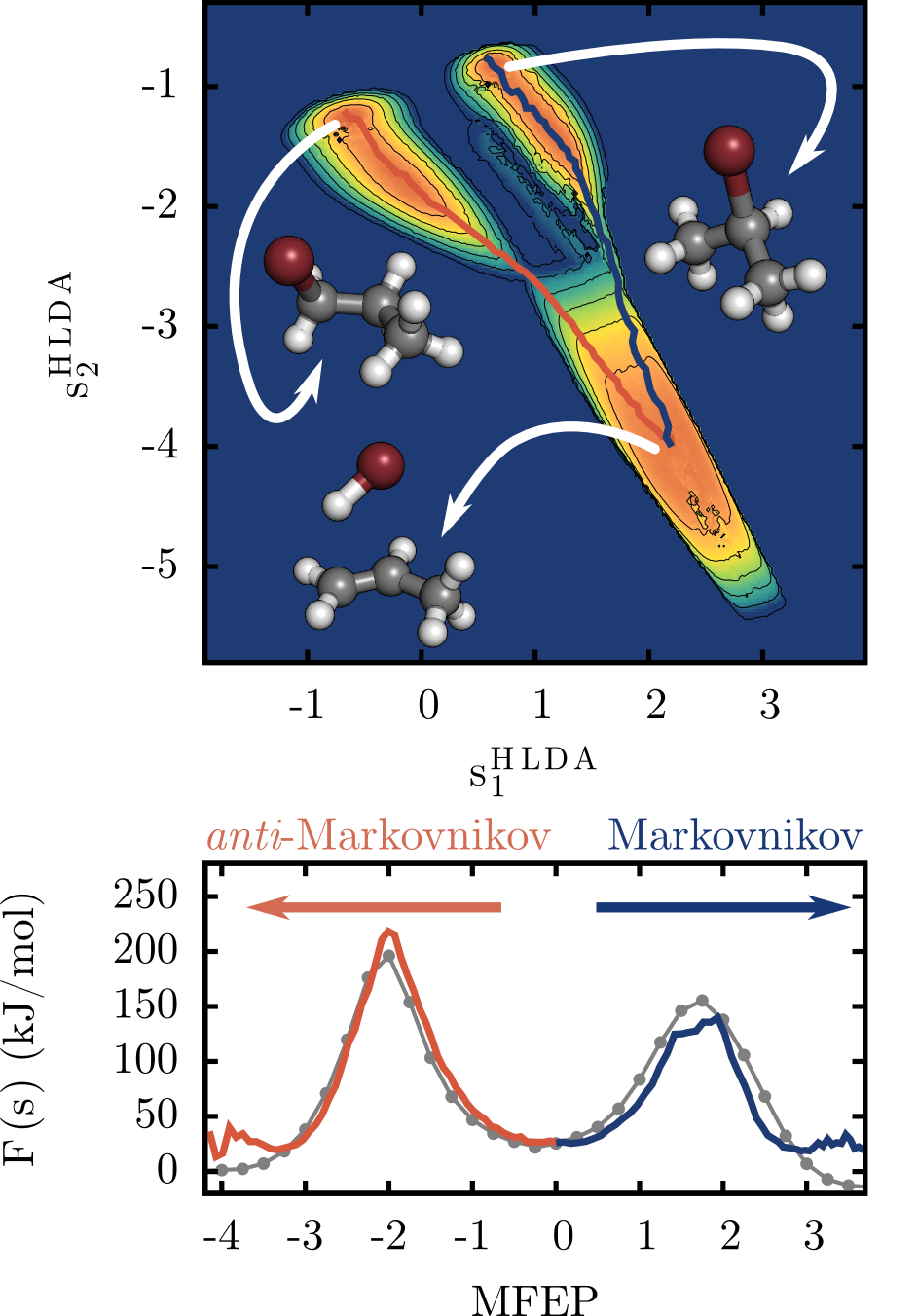}
   	 \caption{Upper panel: free energy surface for the hydrobromination of propene showing the two possible reaction paths and associated reactants and products structures. Lower panel: minimum free energy paths along for the two possible mechanisms of the reaction, namely the high-barrier \emph{anti}-Markovnikov's route (orange line), and the low-barrier Markovnikov's route (blue line).}
    	\label{fig:fes_es}
\end{figure}

Fig. \ref{fig:fes_es} reports in the upper panel the FES of the hydrobromination reaction.
Convergence of the FES with respect to simulation time and relative error bars estimated via bootstrap analysis can be found in the supporting information.
The lower elongated minimum is associated to the reactants state.
This state can evolve into the Markovnikov and anti-Markovnikov product states by crossing two different barriers.
The lower panel of Fig. \ref{fig:fes_es} shows the free energy profile along the MFEP.
Two things are worth noticing in this plot.
Firstly, the difference in thermodynamic stability of the Markovnikov and anti-Markovnikov products is almost negligible.
Secondly, the barriers separating the states differ from each other of about 80 \si{kJ/mol}.
Recalling that the transition probability for a system to transit from one metastable state to another is proportional to the exponential of the free energy barrier it is clear that the nature of the Markovnikov's regioselectivity is almost purely kinetic as deduced from empirical observations.

To further support our conclusion we compare the barriers obtained with metadynamics with the ones calculated by means of a NEB optimization of 60 images for both the Markovnikov and anti-Markovnikov route at 0 \si{K} on the potential energy surface.
The gray dashed dotted line in the lower panel of Fig. \ref{fig:fes_es} represents the energy of the images optimized by means of the NEB algorithm.
It is clear that metadynamics is able to reproduce pretty well the energy landscape (see also Ref.\cite{FRANZEN201646}).
One must keep in mind that metadynamics works on the free energy surface at finite temperature whereas the NEB optimization is applied on the potential energy surface without any account for temperature effects.
This fact is the origin of the slight differences in terms of barrier heights between the two methods.

The large advantage of metadynamics is that all the entropic effects, even the anharmonic ones, are automatically included since the simulation is performed at finite temperature. 
Moreover, NEB optimization may be a rather complicated method if the complexity of the reaction rises with an increasing number of important degrees of freedom.
This happens when the interpolated images between reactants and products lie far from the ideal minimum energy path resulting in a collection of unphysical configurations.
Again, our method does not imply the knowledge of what lies in between reactants and products, as this will be explored automatically by metadynamics directly on the free energy surface, but it rather suggests the proper direction to follow in order to connect them.

\section*{Conclusions}
The method presented in this paper has the potential to become a routine tool in studying chemical reactions.
Its power lies in the combination of a biasing method like Metadynamics with a statistical classification method like harmonic linear discriminant analysis.
The first allows to enhance the fluctuations within the free energy basins in order to favour the transition to other states while the second drives the reaction along the favoured direction towards the desired products.
Moreover, MC-HLDA needs no more information than the one that can be collected from a short unbiased run in the free energy basins of interest.
This fact is fundamental because no assumptions on the possible mechanisms are needed but more importantly no specific reaction pathway is preferentially chosen in order to go from the reactants to the different products.
This approach gives a very clear physical picture of the free energy landscape on which the process takes place.
The accuracy of the derived CVs is such that the estimated barriers along the apparent transition states are close to those obtained by means of standard quantum chemical approaches such as NEB on the potential energy surface at 0 K.
The analysis of the possible reaction paths and relative apparent barriers has allowed us to understand from a free energy perspective the nature of the chirality inversion in the \ce{S_N2} nucleophilic substitution reactions and the kinetic groundings of Markovikov's rules in electrophilic additions.
The method will be soon available as part of the PLUMED2\cite{Tribello2014} program.
We are confident that this method may find a large range of applications, from molecular biology/medicine to heterogeneous and homogeneous catalysis.

\acknowledgement
This research was supported by the European Union Grant No. ERC-2014-AdG-670227/VARMET.
Calculations were carried out on the M\"{o}nch cluster at the Swiss National Supercomputing Center (CSCS).


\bibliography{library}

\newpage
\bigskip
\bigskip

{\centering
\includegraphics[width=0.5\columnwidth]{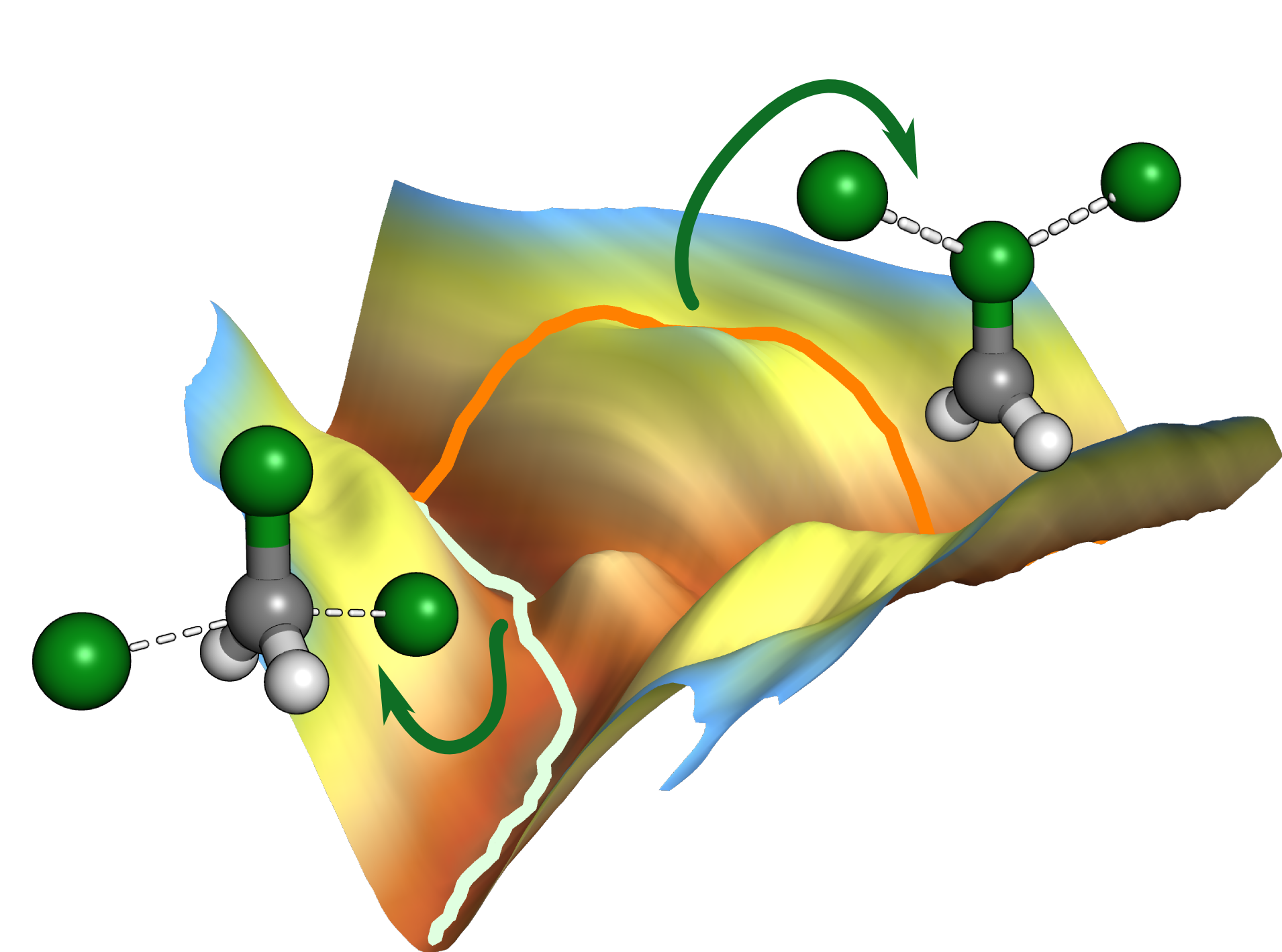}
\par
}

\bigskip
{\sf \small
\noindent
The graphics shows two possible reaction pathways in a \ce{S_N2} reaction that can be explored by means of multiclass linear discriminant analysis metadynamics.
}
\bigskip

\end{document}


\maketitle

\newpage

\section{Computational details}

\subsection*{Nucleophilic substitution of a \ce{Cl} atom in \ce{CH_2 Cl_2}}

Electronic structure calculation:
\begin{itemize}
\item CP2K code\citep{hutter_vandevondele_2014}
\item PM6 Hamiltonian
\item SCF convergence threshold 10E-6 hartree
\item $10\times10\times10$ \si{\angstrom} box
\item Total charge $-1$
\end{itemize}

\bigskip
\noindent
Molecular dynamics calculation:
\begin{itemize}
\item CP2K code\citep{hutter_vandevondele_2014}
\item Temperature 300K
\item NVT Bussi's velocity rescaling thermostat\citep{bussi_parrinello_2007}
\item Timestep 0.5 fs
\end{itemize}

\bigskip
\noindent
MC-HLDA details:
\begin{itemize}
\item Reference run 5 \si{ps}
\item 3 descriptors \ce{C-Cl} distances, see Fig. 1 main text
\item $s_1^{HLDA}$ coefficients = 0.554,-0.811,0.188
\item $s_2^{HLDA}$ coefficients = 0.602,0.237,-0.762
\end{itemize}

\bigskip
\noindent
MC-HLDA metadynamics calculation:
\begin{itemize}
\item Plumed2\citep{tribello_2014} code patched with CP2K
\item Single walker
\item 3 \si{ns} sampling
\item Temperature 300K
\item Well-Tempered MetaD\citep{Barducci2008} with $\gamma=25$
\item 2 \si{kJ/mol} Gaussian height
\item 0.2 \si{\angstrom} Gaussian sigma
\end{itemize}

\bigskip
\noindent
Reference structure: \\ 

{\tt{
\noindent
C          5.09459        4.91798        4.93038 \\
Cl          5.00420        4.99636        2.06235 \\ 
Cl         3.82940        6.01611        4.33492 \\
H          4.89620        3.89061        4.56146 \\ 
H          6.08372        5.25871        4.56148 \\ 
Cl         5.09189        4.92023        6.86383 \\
}}

\newpage
\noindent
FES convergence: \\ 
\begin{figure}[b!]
	\includegraphics[width=.9\columnwidth]{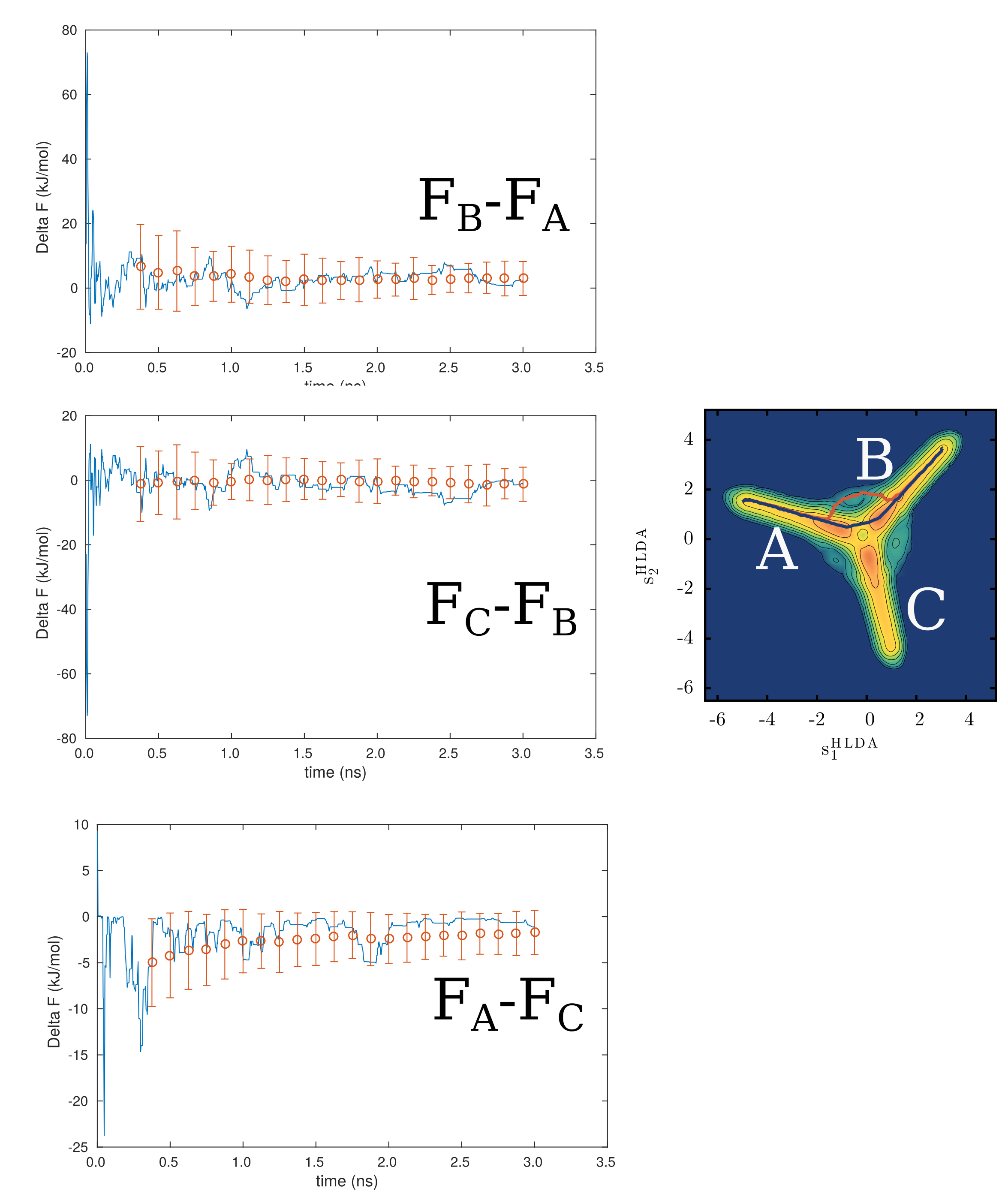}
    	\label{fig:neb}
\end{figure}

\newpage
\subsection*{Electrophilic addition of \ce{HBr} to propene}

Electronic structure calculation:
\begin{itemize}
\item CP2K code\citep{hutter_vandevondele_2014}
\item PM6 Hamiltonian
\item SCF convergence threshold 10E-7 hartree
\item $10\times10\times10$ \si{\angstrom} box
\end{itemize}

\bigskip
\noindent
Molecular dynamics calculation:
\begin{itemize}
\item CP2K code\citep{hutter_vandevondele_2014}
\item Temperature 300K
\item NVT Bussi's velocity rescaling thermostat\citep{bussi_parrinello_2007}
\item Timestep 1.0 fs
\end{itemize}

\bigskip
\noindent
MC-HLDA details:
\begin{itemize}
\item Reference run 85 \si{ps}
\item 5 descriptors, distances, see Fig. 4 main text
\item $s_1^{HLDA}$ coefficients = -0.523,0.726,0.401,-0.029,-0.193
\item $s_2^{HLDA}$ coefficients = 0.573,-0.339,-0.183,-0.636,-0.345
\end{itemize}

\bigskip
\noindent
MC-HLDA metadynamics calculation:
\begin{itemize}
\item Plumed2\citep{tribello_2014} code patched with CP2K
\item 3 multiple-walkers
\item 2 \si{ns} sampling per walker
\item Temperature 300K
\item Well-Tempered MetaD\citep{Barducci2008} with $\gamma=50$
\item 1 \si{kJ/mol} Gaussian height
\item 0.2 \si{\angstrom} Gaussian sigma
\end{itemize}

\bigskip
\noindent
Reference structure: \\ 

{\tt{
\noindent
C        -1.777893        0.103815        0.254504\\
C        -0.440198        0.405579        0.380877\\
H        -1.946432       -0.713641       -0.163503\\
H        -2.556681        0.529777        0.591753\\
C         0.790177       -0.266775       -0.193057\\
H        -0.126084        1.089115        1.184146\\
H         0.556427       -1.152008       -0.809697\\
H         1.517651       -0.609212        0.509388\\
H         1.274498        0.412571       -0.908895\\
Br       -1.131277        2.976166       -0.082215\\
H        -2.104167        3.470361        0.850613\\
}}

\newpage
\noindent
FES convergence: \\ 
\begin{figure}[b!]
	\includegraphics[width=.9\columnwidth]{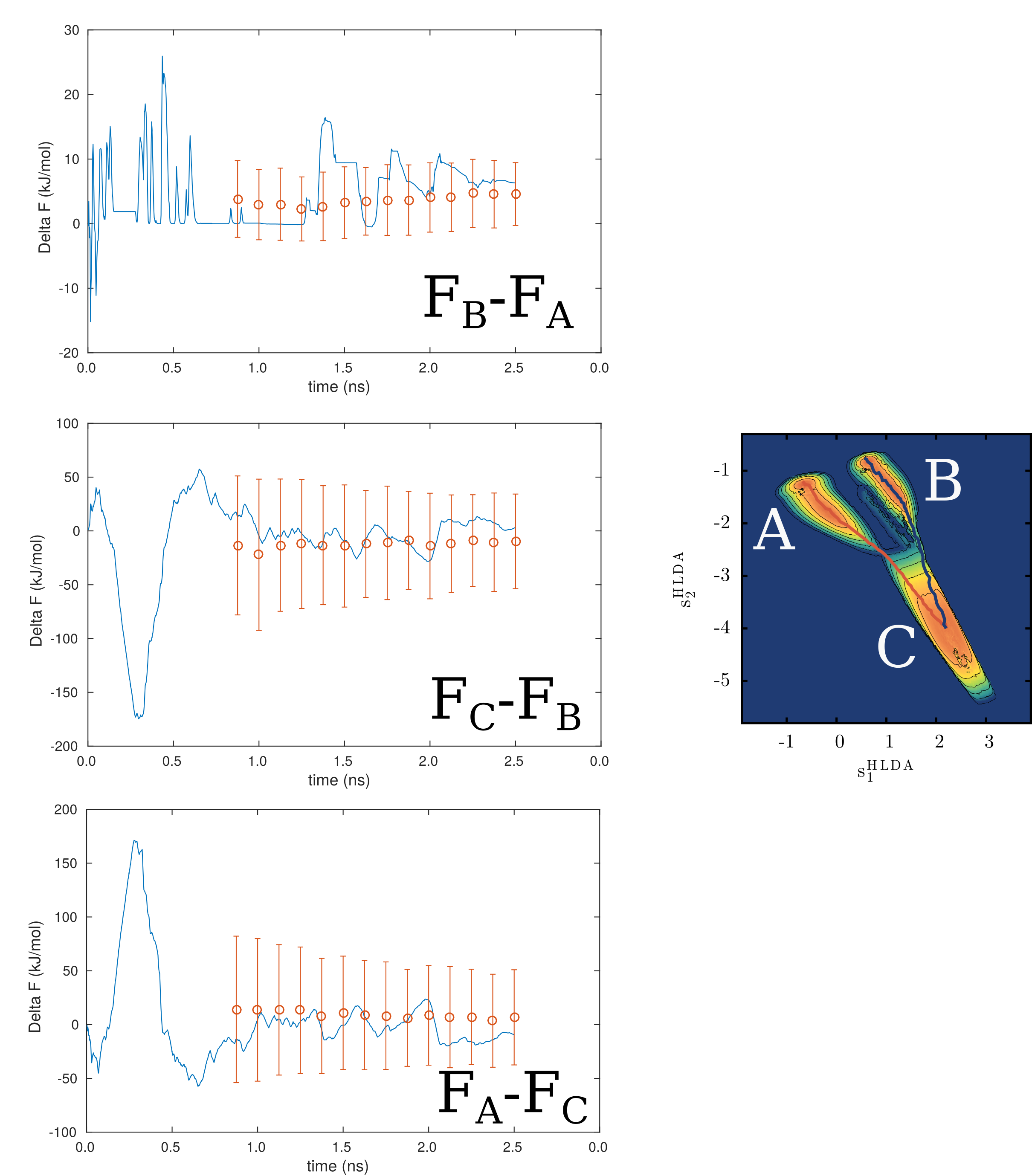}
    	\label{fig:neb}
\end{figure}

\newpage
\subsection*{NEB calculations at 0 K}

NEB details:
\begin{itemize}
\item CP2K code
\item 60 images per reaction path (Markovnikov and anti-Markovnikov)
\item Algorithm Climbing Image NEB
\item $K_{spring}=0.02$
\item Optimization algorithm: DIIS
\end{itemize}

\bigskip
\noindent 
Fig. \ref{fig:neb} reportes the total energy profiles of the NEB calculations.
In Fig. 5 of the main text the NEB images distance has been adjusted and rescaled in order to match the position of the metadynamics peaks and shifted to the reactants energy.
This does not imply any loss of generality as long as what we are comparing are the barrier heights.

\begin{figure}
	\includegraphics[width=1.\columnwidth]{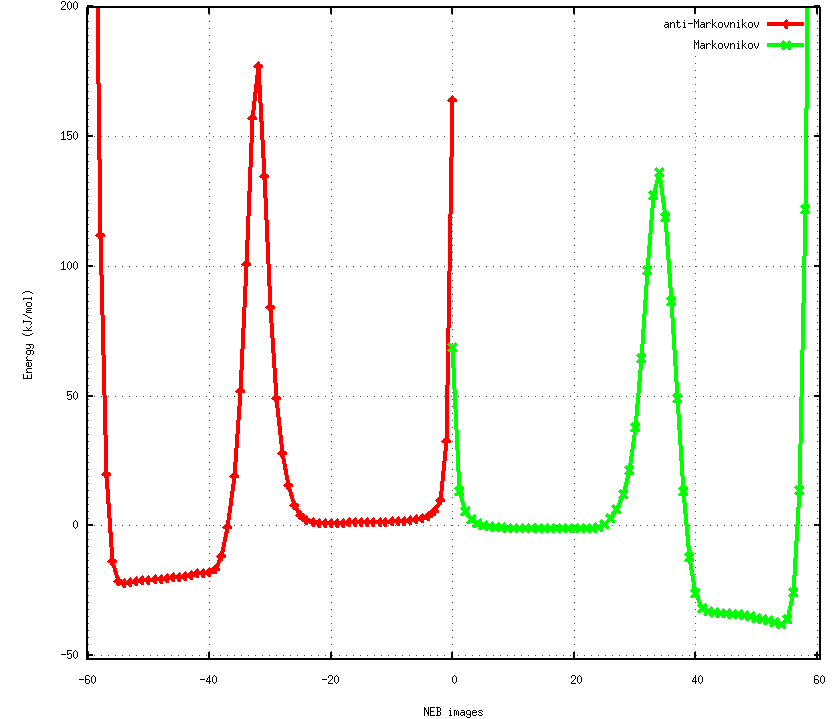}
   	 \caption{NEB energy profiles for the independent Markovnikov and anti-Markovnikov reaction paths.}
    	\label{fig:neb}
\end{figure}

\newpage
\subsection*{MC-HLDA code availability}

The MC-HLDA code providing the coefficients of the linear combination has been first developed using Python by GM. Piccini.
A development version to be used in PLUMED2 has been recently developed by Dr. Y. I. Yang and is available at the following URL:

\bigskip
\noindent
{\tt{
https://github.com/helloyesterday/plumed2\_HLDA
}}

\bigskip
\noindent
We kindly acknowledge Dr. Y. I. Yang for implementing this first version of the code.

\newpage
\bibliography{library}
\bibliographystyle{plain}